\documentclass[reprint,aps,prb,twocolumn,groupedaddress]{revtex4-2}
\usepackage{graphicx}
\usepackage{here}
\usepackage{amsmath}
\usepackage{braket}
\usepackage{amsfonts}
\usepackage{subcaption}
\usepackage{blkarray}
\usepackage{makecell}
\usepackage{ragged2e}
\usepackage{bm}
\usepackage{xcolor}

\begin{document}

\title{
Effective phonon models based on symmetry-adapted multipole basis \\
-- Hidden chiral phonon angular momentum splitting in ferroaxial systems
}

\author{Yu Xie, Rikuto Oiwa, and Satoru Hayami}
\affiliation{Graduate School of Science, Hokkaido University, Sapporo 060-0810, Japan} %\\

\begin{abstract}
We propose a symmetry-based framework for constructing effective harmonic phonon models using a symmetry-adapted multipole basis.
By decomposing the force-constant matrix into bond-centered electric multipoles, we identify the minimal microscopic ingredients responsible for phonon angular-momentum splitting.
Applying this framework to a minimal zigzag-chain model, we {show} that ferroaxial order {gives rise to} a hidden sublattice-resolved chiral phonon{s}, while an additional polar contribution {leads to} finite global chirality.
Our results provide a unified symmetry-based description of hidden and emergent phonon phenomena and suggest a route to control phonon properties via electronic orderings and external fields.
\end{abstract}

\maketitle

%%%%%%%%%%%%%
%%%  Introduction  %%%
%%%%%%%%%%%%%

Chirality is a geometric property defined in three dimensions by the simultaneous breaking of spatial inversion $\mathcal{P}$ and all mirror symmetries, and is characterized by a time-reversal $\mathcal{T}$-even pseudoscalar~\cite{Barron2013}.
In phononic systems, chiral phonons~\cite{HanyuEtAl2018Science,ChenEtAl2022NanoLetters,BousquetEtAl2025JPCM,KatoKishine2023JPSJ,WangEtAl2024NanoLett,juraschekEtAl2025NaturePhys} carrying angular momentum (AM) along the propagation vector $\bm{q}$ have attracted significant attention~\cite{ZhangNiu2014PRL,ZhangNiu2015PRL,TatsumiEtAl2018PRB,ZhangEtAl2022PRR,Komiyama2022PRB,ZhangEtAl2025NatPhys}, as their energy dispersions split according to the phonon AM and have been experimentally detected using Raman and X-ray spectroscopies~\cite{PineDresselhaus1969PhysRev,IshitoEtAl2023NatPhys,IshitoEtAl2023Chirality,OishiEtAl2024PRB,UedaEtAl2023Nature}.
Chiral phonons give rise to various phenomena, including temperature-gradient-induced phonon AM~\cite{HamadaEtAl2018PRL}, chiral-phonon-induced electric current~\cite{YaoMurakami2022PRB} and spin polarization~\cite{Fransson2023PRR}, and electron--chiral-phonon coupling~\cite{TateishiEtAl2025JPSJ}.
They have also been discussed in connection with the selective excitation of phonon AM~\cite{OheEtAl2024PRL} and chirality-induced spin selectivity~\cite{KatoEtAl2022PRB}.

On the other hand, ferroaxiality represents a distinct form of geometric order characterized by a $\mathcal{T}$-even axial vector while preserving $\mathcal{P}$ symmetry~\cite{JinEtAl2020NatPhys,HlinkaEtAl2014PRL,HlinkaEtAl2016PRL}.
Following the experimental identification of ferroaxial transitions in materials such as RbFe(MoO$_4$)$_2$~\cite{JinEtAl2020NatPhys,OwenEtAl2021PRB}, it has been reported in a wide range of compounds, including NiTiO$_3$~\cite{HayashidaEtAl2020NatCommun,HayashidaEtAl2021PRMater, YokotaEtAl2022npjQuantumMater,FangEtAl2023JAmChemSoc,GuoEtAl2023PRB}, Ca$_5$Ir$_3$O$_{12}$~\cite{Hasegawa_doi:10.7566/JPSJ.89.054602, hanate2021first, hanate2023space, hayami2023cluster}, and K$_2$Zr(PO$_4$)$_2$~\cite{YamagishiEtAl2023ChemMater, XieOiwaHayami2025arxiv}.
In addition, chirality can be induced in such systems through coupling to polar degrees of freedom under an external electric field~\cite{HayashidaEtAl2023PNAS, hayami2025chirality}, providing a route to control chiral responses in intrinsically nonchiral crystals. 
Recent studies have further shown circular-dichroic Raman responses even in centrosymmetric ferroaxial crystals without structural chirality~\cite{Watanabe2025PRB,KusunoEtAl2026PRL,Watanabe2026arxiv,Suganuma2026arXiv}. 
These findings position ferroaxial systems as a versatile platform for chirality control in phononic and optical responses. 
However, the influence of ferroaxiality on phonon dispersions remains largely unexplored, and the microscopic origin of phonon AM splitting is still unclear.
While hidden spin polarization has been identified~\cite{BhowalSpaldin2024PRR}, the existence and origin of hidden phonon 
{AM} remain open questions.

From a theoretical perspective, phonon AM splitting has been clarified in intrinsically chiral crystals.
In monoaxial systems, only optical modes show linear-in-$q$ splitting, whereas in cubic systems, it appears only in optical triplets near the $\Gamma$ point~\cite{TsunetsuguKusunose2023JPSJ, TsunetsuguKusunose2026JPSJ}.
By contrast, in ferroaxial systems, the microscopic origin remains unclear.
In particular, it is unknown which force-constant components are responsible for (i) hidden phonon AM splitting and (ii) the emergence of global chirality when polarity is introduced.
Furthermore, a unified microscopic description treating intrinsic chirality, ferroaxiality, and polarity on an equal footing is still lacking.

In this study, we construct harmonic phonon models based on a symmetry-adapted multipole basis (SAMB)~\cite{HayamiEtAl2018PRB,KusunoseOiwaHayami2020JPSJ,HayamiKusunose2024JPSJ,OiwaKusunose2022PRL,KusunoseOiwaHayami2023PRB,IndaEtAl2024JCP,OiwaKusunose2025PRR, oiwa2025symmetry}.
By expanding the positive semidefinite force constant matrix in terms of SAMB and imposing {both} symmetry constraints and the acoustic sum rule {(ASR)}, we extract the minimal microscopic degrees of freedom associated with polarity and ferroaxiality.
Applying this framework to a zigzag-chain model, we reveal the origin of hidden phonon AM in ferroaxial systems and {clarify how it evolves into a finite} 
global chirality.

%%%%%%%%%%%%%
%%%      Method     %%%
%%%%%%%%%%%%%

Let us begin with a general formulation of effective phonon models. 
Within the harmonic approximation, the potential energy $U$ is expanded up to second order in lattice displacements.
Introducing the mass-weighted displacements $u_{i}^{\alpha}(\bm{R}, t) = \sqrt{m_i}\, x_{i}^{\alpha}(\bm{R}, t)$, where $x_{i}^{\alpha}(\bm{R}, t)$ denotes the $i$th-sublattice atomic displacements for the $\alpha \in {x,y,z}$ direction ($\bm{R}$ is a Bravais lattice vector and $t$ is the time) and $m_{i}$ is the atomic mass, the energy is written as
\begin{align}
&U = \frac{1}{2}\sum_{\bm{R},\bm{R}'} \sum_{i,j} \sum_{\alpha,\beta}u_{i}^{\alpha}(\bm{R}, t) \Phi_{ij}^{\alpha\beta}(\bm{R}-\bm{R}')u_{j}^{\beta}(\bm{R}', t),
\label{eq_harmonic_pot}
\\
&\Phi_{ij}^{\alpha\beta}(\bm{R}-\bm{R}')  = \frac{\partial^{2} U}{\partial u_{i}^{\alpha}(\bm{R},t) \partial u_{j}^{\beta}(\bm{R}',t)},
\label{force_const}
\end{align}
where $\Phi_{ij}^{\alpha\beta}(\bm{R}-\bm{R}')$ is the mass-weighted force constants.
The corresponding dynamical matrix is given by
\begin{align}
D_{ij}^{\alpha\beta}(\bm{q}) = \frac{1}{N} \sum_{\bm{R}} \Phi_{ij}^{\alpha\beta}(\bm{R}) e^{-i \bm{q} \cdot (\bm{R} + \bm{r}_{i} - \bm{r}_{j})},
\label{eq_Dynamical_matrix}
\end{align}
where $N$ is the number of unit cells, and $\bm{r}_i$ is the position of {$i$th-}sublattice 
within the unit cell.
The eigenvalues $\lambda_{n\bm{q}}$ and eigenvectors $\bm{\epsilon}_{n\bm{q}}$ of $\hat{D}(\bm{q})$ are obtained as $\hat{U}_{\bm{q}}^{\dagger} \hat{D}(\bm{q}) \hat{U}_{\bm{q}} = \mathrm{diag}(\lambda_{1\bm{q}}, \lambda_{2\bm{q}}, \cdots, \lambda_{M\bm{q}})$ and $\hat{U}_{\bm{q}}  = [\bm{\epsilon}_{1\bm{q}}, \bm{\epsilon}_{2\bm{q}}, \cdots, \bm{\epsilon}_{M\bm{q}}]$, from which the phonon frequencies are given by $\omega_{n\bm{q}} = \sqrt{\lambda_{n\bm{q}}}$.
The displacements $\bm{u}_{i}(\bm{R}, t)$ can then be expressed as
\begin{align}
\bm{u}_{i}(\bm{R}, t) = \frac{1}{2}\frac{1}{\sqrt{N}} \sum_{n\bm{q}} \bm{\epsilon}_{n\bm{q}}^{*} e^{i [\bm{q} \cdot (\bm{R} + \bm{r}_{i}) - \omega_{n\bm{q}} t]}  + \mathrm{c.c.}
\label{eq_u_FT}
\end{align}

By introducing a composite index $a \equiv (i,\bm{R})$, the potential energy $U$ can be rewritten as
\begin{align}
U = \frac{1}{2} \sum_{ab} \bm{u}_a^{\rm T}  \hat{\Phi}_{ab} \bm{u}_b,
\end{align}
where the explicit time dependence in $\bm{u}$ is omitted for simplicity.
For a physically valid harmonic potential, the force-constant matrix $\hat{\Phi}$ must satisfy the following three constraints.
\begin{enumerate}
\item Hessian symmetry:
\begin{align}
\hat{\Phi}_{ab} = \hat{\Phi}_{ba}^{\rm T}.
\label{eq_hessian}
\end{align}
\item 
{ASR}:
\begin{align}
\sum_{b} \hat{\Phi}_{ab} = 0,
\quad
\hat{\Phi}_{aa} =  -\sum_{b \neq a}
\hat{\Phi}_{ab}.
\label{eq_asr}
\end{align}
This condition follows from translational invariance and guarantees the existence of three acoustic phonon modes with zero frequency at the $\Gamma$ point.
The on-site matrix elements $\hat{\Phi}_{aa}$ are fully determined by the inter-site matrix elements $\hat{\Phi}_{ab}$ $(a \neq b)$ through this constraint.
\item Positive semidefiniteness:
\begin{align}
U(\{\bm{u}\}) \ge 0,
\label{eq_positive_definite}
\end{align}
with equality 
only for uniform translations{, ensuring mechanical stability.} 
Equivalently, {$\hat{\Phi}$ is positive semidefinite and} 
$\hat{D}(\bm{q})$ has non-negative eigenvalues, $\omega^2(\bm{q}) \ge 0$.
\end{enumerate}

We restrict ourselves to an effective harmonic model in which the energy is expressed as a sum of bond contributions depending only on relative displacements:
\begin{align}
U = -\frac{1}{4} \sum_{a\neq b}
\Delta \bm{u}_{ab}^{\rm T}
\hat{\Phi}_{ab}
\Delta \bm{u}_{ab},
\qquad
\Delta \bm{u}_{ab} = \bm{u}_a - \bm{u}_b
{.}
\label{eq_bond_stiffness}
\end{align}
Such bond-based formulations are widely used in lattice dynamics 
{for their} transparent description of 
bond-dependent force constants and are 
{compatibility} with the mechanical stability condition (2).
They have also 
{applied to} chiral phonon systems~\cite{TsunetsuguKusunose2023JPSJ}.
In this representation, only the symmetric part $\hat{\Phi}_{ab}^{\rm (S)} = \hat{\Phi}_{ab}^{\rm (S) T}$ contributes to the energy, since $\Delta \bm{u}_{ab}^{\rm T} \hat{\Phi}_{ab}^{\rm (A)} \Delta \bm{u}_{ab} = 0$ for any antisymmetric matrix $\hat{\Phi}_{ab}^{\rm (A)} = -\hat{\Phi}_{ab}^{\rm (A) T}$.
Accordingly, we take $\hat{\Phi}_{ab}$ to be symmetric without loss of generality in this work:
\begin{align}
\hat{\Phi}_{ab} = \hat{\Phi}_{ab}^{\rm T}, \quad (a\neq b).
\label{eq_symmetric_xy}
\end{align}
Combining this with the Hessian symmetry in Eq.~(\ref{eq_hessian}), we obtain
\begin{align}
\hat{\Phi}_{ab} = \hat{\Phi}_{ba}, \quad (a\neq b).
\label{eq_symmetric_ab}
\end{align}

\begin{figure}[t]
\includegraphics[width=1.0\linewidth]{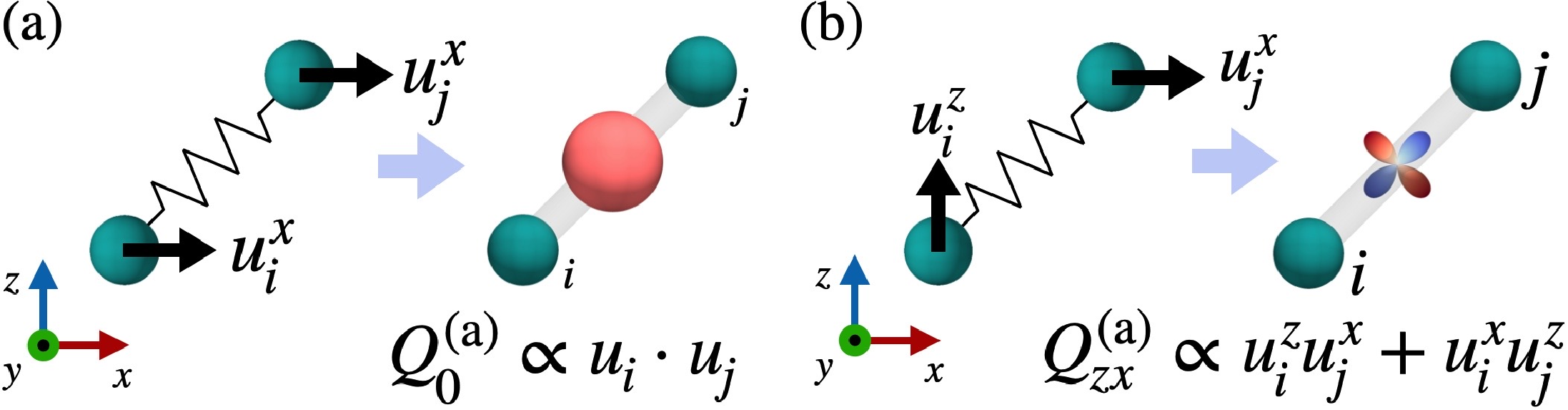}
\caption{\justifying
(a) Schematic illustration of the 
{force} constant for the {diagonal} isotropic 
{interaction}, {corresponding to the electric monopole} $Q_0^{\rm (a)}$ {at {the} bond center}, and (b) that for the {off-diagonal} anisotropic 
{interaction}, {corresponding to the electric quadrupole} $Q_{zx}^{\rm (a)}$ {at {the} bond center}.
}
\label{fig_spring_samb}
\end{figure}

{
Using the symmetry constraints on the force constants given by Eqs.~(\ref{eq_hessian}), (\ref{eq_asr}), (\ref{eq_symmetric_xy}), and (\ref{eq_symmetric_ab}), we show a method to construct effective {harmonic} phonon models 
based on 
{SAMB}~\cite{HayamiEtAl2018PRB,KusunoseOiwaHayami2020JPSJ,HayamiKusunose2024JPSJ,OiwaKusunose2022PRL,KusunoseOiwaHayami2023PRB,IndaEtAl2024JCP,OiwaKusunose2025PRR, oiwa2025symmetry}.
Since 
{each displacement component $u^{\alpha}$ transforms in the same way} as the electronic $p_{\alpha}$ orbital, the $3\times 3$ symmetric force constant matrix $\hat{\Phi}_{ab} = \hat{\Phi}_{ab}^{\rm T}$ for 
{a given} pair $(a,b)$ can be described by using 
{six electric-type atomic} SAMBs:
}
 \begin{align}
 \begin{split}
 &
 \hat{Q}_0^{\rm (a)}
 =
 \frac{1}{\sqrt{3}}
 \begin{pmatrix}
 1&0&0\\
 0&1&0\\
 0&0&1\\
 \end{pmatrix}, \,
\hat{Q}_{u}^{\rm (a)}
 =
\frac{1}{\sqrt{6}}
 \begin{pmatrix}
 -1&0&0\\
 0&-1&0\\
 0&0&2\\
 \end{pmatrix}, \\
&
 \hat{Q}^{\rm (a)}
 _{v}=
\frac{1}{\sqrt{2}}
 \begin{pmatrix}
 1&0&0\\
 0&-1&0\\
 0&0&0\\
 \end{pmatrix}, \,
 \hat{Q}_{yz}^{\rm (a)}
 =
 \frac{1}{\sqrt{2}}
 \begin{pmatrix}
 0&0&0\\
 0&0&1\\
 0&1&0\\
 \end{pmatrix}, \\
& \hat{Q}_{zx}^{\rm (a)}
 =
 \frac{1}{\sqrt{2}}
 \begin{pmatrix}
 0&0&1\\
 0&0&0\\
 1&0&0\\
 \end{pmatrix}, \,
 \hat{Q}_{xy}^{\rm (a)}
 =
 \frac{1}{\sqrt{2}}
 \begin{pmatrix}
 0&1&0\\
 1&0&0\\
 0&0&0\\
 \end{pmatrix},
 \end{split}
 \label{eq_Qxyzx}
 \end{align}
where $\hat{Q}^{\rm (a)}_{0}$ is the rank-0 electric monopole and $(\hat{Q}^{\rm (a)}_{u}, \hat{Q}^{\rm (a)}_{v}, \hat{Q}^{\rm (a)}_{yz}, \hat{Q}^{\rm (a)}_{zx}, \hat{Q}^{\rm (a)}_{xy})$ are the rank-2 electric quadrupoles.

As illustrated in Fig.~\ref{fig_spring_samb}(a), $ \hat{Q}_0^{\rm (a)}$ corresponds to force constants that couple $x$--$x$, $y$--$y$, and $z$--$z$ displacements, representing a diagonal isotropic interaction.
In contrast, $\hat{Q}^{\rm (a)}_{zx}$ corresponds to a force constant that couples $z$--$x$ displacements, giving rise to an off-diagonal anisotropic interaction, as shown in Fig.~\ref{fig_spring_samb}(b).
Although these atomic multipoles are even under $\mathcal{P}$, they can serve as the microscopic origin of phonon AM splitting when embedded in appropriate lattice symmetry~\cite{Hayami_PhysRevLett.122.147602, ishitobi2026purely}. 

Meanwhile, the remaining three antisymmetric components of the force-constant matrix under the exchange $\alpha \leftrightarrow \beta$ correspond to the atomic magnetic dipoles, given by $[M_{\gamma}^{\rm (a)}]^{\alpha\beta} = -i \epsilon_{\alpha\beta\gamma} / \sqrt{2}$, which do not con{t}ribute to the harmonic potential $U$ given by Eq.~(\ref{eq_bond_stiffness}) due to the requirement in Eq.~(\ref{eq_symmetric_xy}). 
Note that $\bm{M}^{\rm (a)}$ is directly related to the phonon AM as~\cite{ZhangNiu2014PRL}
\begin{align}
\bm{L}^{\rm (ph)} = \sqrt{2} \bm{M}^{\rm (a)}. 
\label{eq_Lph}
\end{align}
The phonon AM splitting discussed below can be described by couplings between odd-order wave-vector components $q_{l}$ and $\bm{L}^{\rm (ph)}$.

Furthermore, we introduce site- and bond-cluster SAMBs, denoted as $[\hat{Y}_{l}^{\rm (s/b)}]_{ab} = [\hat{Y}_{l}^{\rm (s/b)}]_{ba}$ $(Y=Q/T)$, where rank $l$ is the rank of multipole.
These quantities encode the spatial distribution and relative weight of multipoles at each site or bond, with $Q$ and $T$ representing the electric and magnetic toroidal multipoles, respectively.
When the multipole is defined at the bond center between two sites, it is referred to as bond-cluster SAMBs~\cite{KusunoseOiwaHayami2023PRB}.

As a simple example, consider the two-site cluster shown in Fig.~\ref{fig_spring_samb}.
The site-cluster SAMBs consist of the electric monopole and the electric dipole, whose $2\times 2$ matrix representation in the sublattice basis (A, B) are given by
 \begin{align}
 \begin{split}
 &
 \hat{Q}_0^{\rm (s)}
 =
 \frac{1}{\sqrt{2}}
 \begin{pmatrix}
 1&0\\
 0&1\\
 \end{pmatrix}, \,
\hat{Q}_{\parallel}^{\rm (s)}
 =
\frac{1}{\sqrt{2}}
 \begin{pmatrix}
 1&0\\
 0&-1\\
 \end{pmatrix}, \\
\end{split}
 \label{eq_Qsite}
 \end{align}
where $\parallel$ means the bond-direction component between two sites.
Similarly, the bond-cluster SAMBs are described by the electric monopole and magnetic toroidal dipole, which are expressed as
 \begin{align}
 \begin{split}
 &
 \hat{Q}_0^{\rm (b)}
 =
 \frac{1}{\sqrt{2}}
 \begin{pmatrix}
 0&1\\
 1&0\\
 \end{pmatrix}, \,
\hat{T}_{\parallel}^{\rm (b)}
 =
\frac{1}{\sqrt{2}}
 \begin{pmatrix}
 0&i\\
 -i&0\\
 \end{pmatrix}. \\
\end{split}
 \label{eq_Qsite}
 \end{align}
Because of Eq.~(\ref{eq_symmetric_ab}), the bond-cluster magnetic toroidal SAMBs, which are antisymmetric for $a \leftrightarrow b$, $[\hat{T}_{l}^{\rm (b)}]_{ab} = -[\hat{T}_{l}^{\rm (b)}]_{ba}$, cannot be incorporated into the harmonic potential $U$.

The force constant matrix $[\hat{\Phi}]_{ab}^{\alpha\beta} = \delta_{ab} [\hat{\Phi}^{\rm (on)}]_{aa}^{\alpha\beta} + (1 -  \delta_{ab}) [\hat{\Phi}^{\rm (inter)}]_{ab}^{\alpha\beta}$ can be systematically constructed using on-site and inter-site SAMBs, denoted as $\hat{\mathbb{Z}}_{j}^{\rm (on)}$ and $\hat{\mathbb{Z}}_{j}^{\rm (inter)}$, respectively.
These basis functions are expressed as linear combinations of direct products between the six atomic electric SAMBs $\hat{Q}_{l_{1}\xi_{1}}^{\rm (a)}$ and the site- or bond-cluster electric SAMBs $\hat{Q}_{l_{2}\xi_{2}}^{\rm (s/b)}$:
\begin{align}
&[\hat{\mathbb{Z}}_{j}^{\rm (on)}]_{aa}^{\alpha\beta} = \sum_{\xi_{1}\xi_{2}} C_{l\xi}^{l_{1}\xi_{1}, l_{2}\xi_{2}}(Z) [\hat{Q}_{l_{1}\xi_{1}}^{\rm (a)}]^{\alpha\beta} \otimes [\hat{Q}_{l_{2}\xi_{2}}^{\rm (s)}]_{aa},
\\
&[\hat{\mathbb{Z}}_{j}^{\rm (inter)}]_{ab}^{\alpha\beta} = \sum_{\xi_{1}\xi_{2}} C_{l\xi}^{l_{1}\xi_{1}, l_{2}\xi_{2}}(Z) [\hat{Q}_{l_{1}\xi_{1}}^{\rm (a)}]^{\alpha\beta} \otimes [\hat{Q}_{l_{2}\xi_{2}}^{\rm (b)}]_{ab},\quad (a \neq b),
\end{align}
where $\xi = (\Gamma, n, \gamma)$, and $\Gamma$ and $\gamma$ denote the irreducible representation (irrep.) and its component, respectively, and the label $n$ represents the multiplicity to distinguish independent SAMBs belonging to the same irrep.
Here, we introduce the composite index $j = (Z, l, \Gamma, n, \gamma)$, where $Z = Q/G$, and $\mathbb{Q}_{l\xi}$ and $\mathbb{G}_{l\xi}$ correspond to the electric and the electric toroidal multipoles with $(\mathcal{P}, \mathcal{T}) = [(-1)^{l}, +1]$ and $ [(-1)^{l+1}, +1]$, respectively.
The coefficients $C_{l\xi}^{l_{1}\xi_{1}, l_{2}\xi_{2}}(Z)$ are orthonormal Clebsch-Gordan (CG) coefficients arising from the reduction to the irrep. $\Gamma$ from the direct product of $\Gamma_{1}$ and $\Gamma_{2}$~\cite{KusunoseOiwaHayami2023PRB}. 
The full set of SAMBs $\{\mathbb{Z}_{j}\}$ can be generated automatically using the Python library, \texttt{MultiPie}~\cite{KusunoseOiwaHayami2023PRB}.

Utilizing the completeness and orthonormality of the SAMBs $\{\hat{\mathbb{Z}}_{j}\}$, the force-constant matrix $\hat{\Phi}$ can be systematically expanded as a linear combination of $\{\hat{\mathbb{Z}}_{j}\}$ as
\begin{align}
\begin{split}
\hat{\Phi}^{\rm (on)} &= \sum_{j} z_{j}^{\rm (on)} \hat{\mathbb{Z}}_{j}^{\rm (on)},
\quad
z_{j}^{\rm (on)} = \mathrm{Tr}\!\left[\hat{\mathbb{Z}}_{j}^{\rm (on)} \hat{\Phi}^{\rm (on)}\right],
\\
\hat{\Phi}^{\rm (inter)} &= \sum_{j} z_{j}^{\rm (inter)} \hat{\mathbb{Z}}_{j}^{\rm (inter)},
\quad
z_{j}^{\rm (inter)} = \mathrm{Tr}\!\left[\hat{\mathbb{Z}}_{j}^{\rm (inter)} \hat{\Phi}^{\rm (inter)}\right],
\end{split}
\label{eq_SymPhononModel}
\end{align}
where the coefficients $\{z_{j}^{\rm (on)}\}$ and $\{z_{j}^{\rm (inter)}\}$ serve as model parameters and are not independent, but are constrained by the ASM in Eq.~(\ref{eq_asr}).
In particular, the on-site coefficients are determined from the inter-site ones as
\begin{align}
z_{i}^{\rm (on)} = \sum_{j} z_{j}^{\rm (inter)} \sum_{a \neq b}
\mathrm{Tr}
\left\{
[\hat{\mathbb{Z}}_{i}^{\rm (on)}]_{aa}
[\hat{\mathbb{Z}}_{j}^{\rm (inter)}]_{ab}
\right\}.
\end{align}
The SAMBs contributing to $\hat{\Phi}$ belong to the totally symmetric irrep. of the point group of the system. 
Symmetry-breaking SAMBs belonging to other irrep. can also be incorporated to analyze the effects of symmetry breaking arising from structural phase transitions or electronic orderings, as discussed below.

%%%%%%%%%%%%%
%%%      Results     %%%
%%%%%%%%%%%%%

\begin{figure}[t]
\includegraphics[width=1\linewidth]{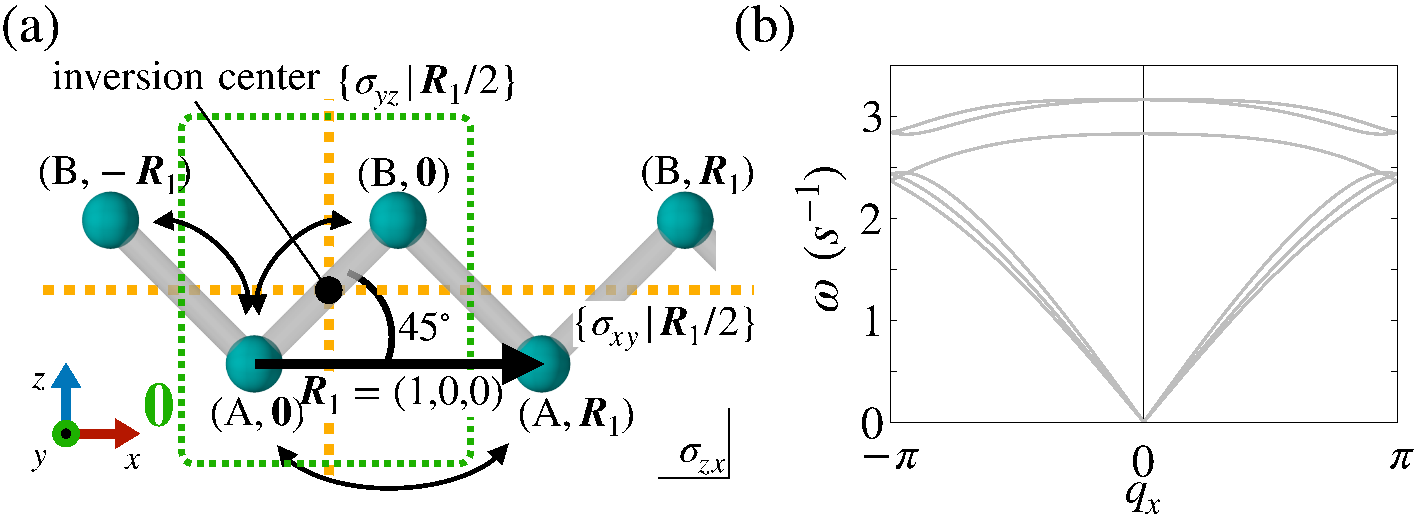}
\caption{\justifying
(a) Schematic illustration of the one-dimensional zigzag chain. 
(b) Phonon energy dispersion of the model.
}
\label{fig_struc}
\end{figure}

We demonstrate our method using a minimal one-dimensional zigzag-chain model composed of A and B sublattices [Fig.~\ref{fig_struc}(a)], belonging to the space group No.~51 with point group {$D_{\rm 2h}$}. 
This model is particularly suitable for our purpose because, despite involving only a simple two-sublattice degree of freedom, it allows for various ordered states arising from the local breaking of 
$\mathcal{P}$ symmetry at each site~\cite{Yanase_JPSJ.83.014703, hayami2016emergent}. 
The potential energy $U{=(1/2)\sum_{ab}^{\rm A,B} \bm{u}_{a}^{\rm T}  \hat{\Phi}_{ab} \bm{u}_{b}}$ is given by
\begin{align}
U = \sum_{n=1}^{{2}} \sum_{ab}^{\rm A,B} k^{(n)} (\Delta \bm{u}_{ab})^{2} + k^{(n)'} {(}\bm{n}_{ab} \cdot \Delta \bm{u}_{ab} {)}^{2},
\label{eq_potential}
\end{align}
where $\bm{n}_{ab}$ is the unit vector along the bond connecting sites $a$ and $b$, and $k^{(n)}$ and $k^{(n)'}$ denote the isotropic and bond-oriented anisotropic force constants {for the $n$th nearest neighbor {(NN)} ($n \le {2}$)}, respectively.
In the following calculations, we set $k^{(1)} = 0.5, k^{(2)} = 0.1${,} 
$k^{(1)'} = 0.25$, {and} $k^{(2)'} = 0.05$.

\begin{figure}[t]
\includegraphics[width=1\linewidth]{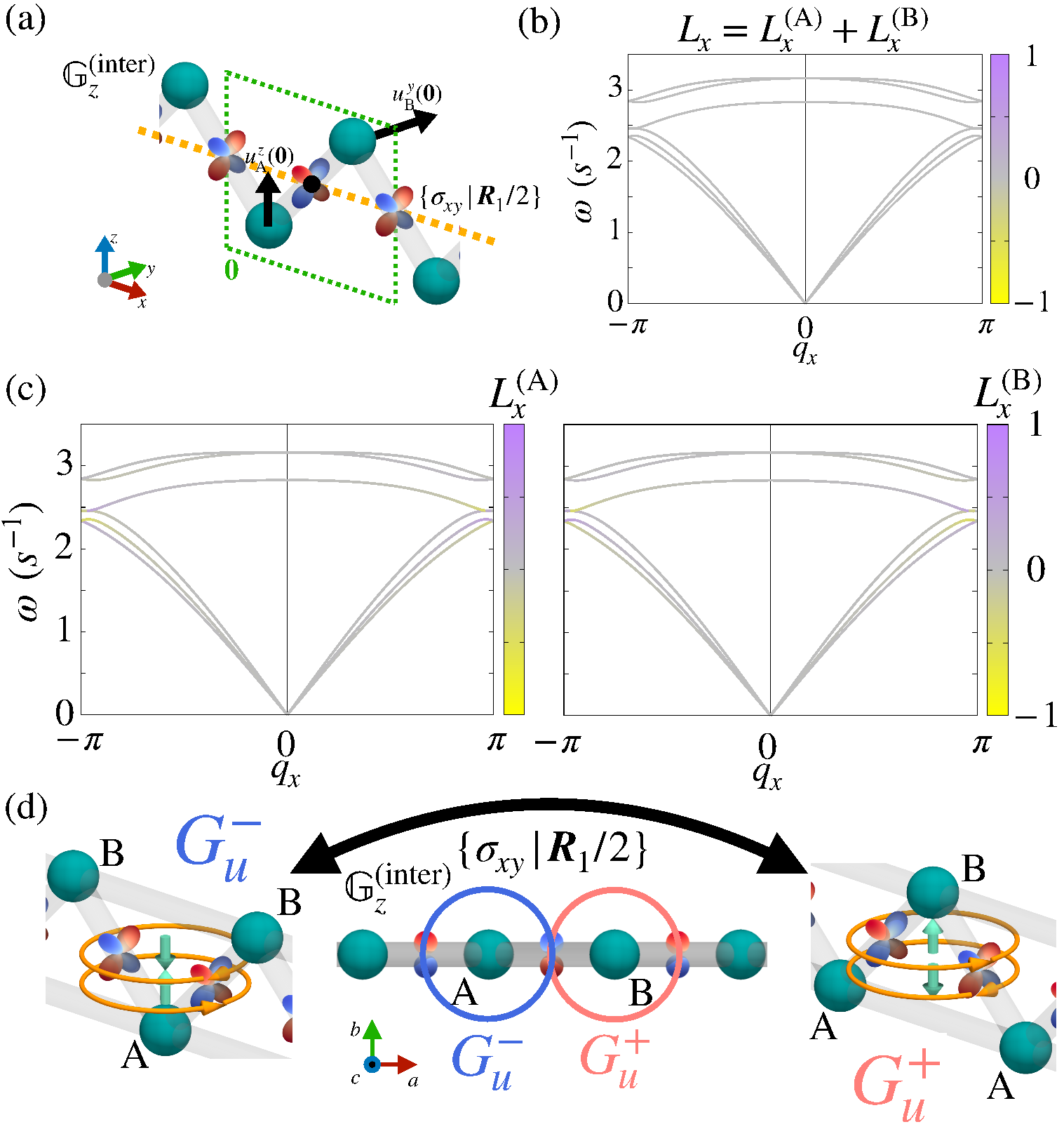}
\caption{\justifying
(a) Schematic representation of the electric toroidal dipole $\hat{\mathbb{G}}_{z}^{\rm (inter)}$ defined in Eq.~(\ref{eq_Gz}), and (b) the corresponding phonon energy dispersion and AM distribution.
(c) Sublattice-resolved phonon AM distribution under ferroaxial order described by $\hat{\mathbb{G}}_{z}^{\rm (inter)}$.
(d) Antiferro-type chiral order of ET quadrupoles $G_u^{\pm}$ corresponding to Fig.~\ref{fig_qz_chiral}(a).
}
\label{fig_gz}
\end{figure}

Using Eq.~(\ref{eq_SymPhononModel}), the $6\times 6$ force-constant matrix $\hat{\Phi}$ can be expressed as a linear combination of SAMBs belonging to th{e} totally symmetric A$_{\rm g}$ irrep. as $\hat{\Phi} = \sum_{j}^{\Gamma_{j} = {\rm A_{g}}} z_{j} \hat{\mathbb{Z}}_{j}$.
One such A$_{\rm g}$ SAMB is given by the electric monopole $\hat{\mathbb{Q}}_{0}^{\rm (inter, A_g)} = \hat{Q}_{zx}^{\rm (a)}\otimes \hat{Q}_{zx}^{\rm (b)}$ where the bond-cluster component satisfies $[\hat{Q}_{zx}^{\rm (b)}]_{\rm AB}(\bm{0}) = [\hat{Q}_{zx}^{\rm (b)}]_{\rm BA}(\bm{0}) = -[\hat{Q}_{zx}^{\rm (b)}]_{\rm AB}(\bm{R}_{1})  = -[\hat{Q}_{zx}^{\rm (b)}]_{\rm BA}(-\bm{R}_{1}) = 1/2$.
This SAMB can be interpreted as the atomic multipole $\hat{Q}_{zx}^{\rm (a)}$ located on the NN bond center between A and B sites.
The resulting phonon energy dispersion obtained from this model is shown in Fig.~\ref{fig_struc}(b).

We first examine the effect of ferroaxial order by introducing a SAMB corresponding to an axial vector along the $z$ direction. 
The relevant term is given by
\begin{align}
\hat{\mathbb{G}}_{z}^{\rm (inter)} = \hat{Q}_{yz}^{\rm (a)} \otimes \hat{Q}_{zx}^{\rm (b)},
\label{eq_Gz}
\end{align}
which belongs to the $\mathrm{B}_{1g}$ irrep. 
This term breaks the mirror symmetries $\sigma_{yz}$ and $\sigma_{zx}$ while preserving $\mathcal{P}$ and the mirror symmetry $\sigma_{xy}$, as illustrated in Fig.~\ref{fig_gz}(a).
Accordingly, the net phonon AM vanishes at each $\bm{q}$ due to the preserved $\mathcal{P}$ and $\mathcal{T}$ symmetries as shown in Fig.~\ref{fig_gz}(b). 

Nevertheless, a nontrivial sublattice-resolved structure emerges. 
As shown in Fig.~\ref{fig_gz}(c), the phonon AM on each sublattice is antisymmetric in $q_x$, satisfying $L_{x}^{\rm (ph),(A,B)}(q_x) = -L_{x}^{\rm (ph),(A,B)}(-q_x)$. 
The A and B contributions have opposite signs and thus cancel in total, resulting in zero net AM. 
This corresponds to a hidden phonon AM, analogous to the sublattice-resolved antisymmetric spin splitting in ferroaxial electronic systems~\cite{BhowalSpaldin2024PRR}. 
As illustrated in Fig.~\ref{fig_gz}(d), it can be viewed as an antiferro-type arrangement of electric toroidal quadrupoles $G_u^{\pm}$, which has the same symmetry as the chirality.

\begin{figure}[t]
\includegraphics[width=1\linewidth]{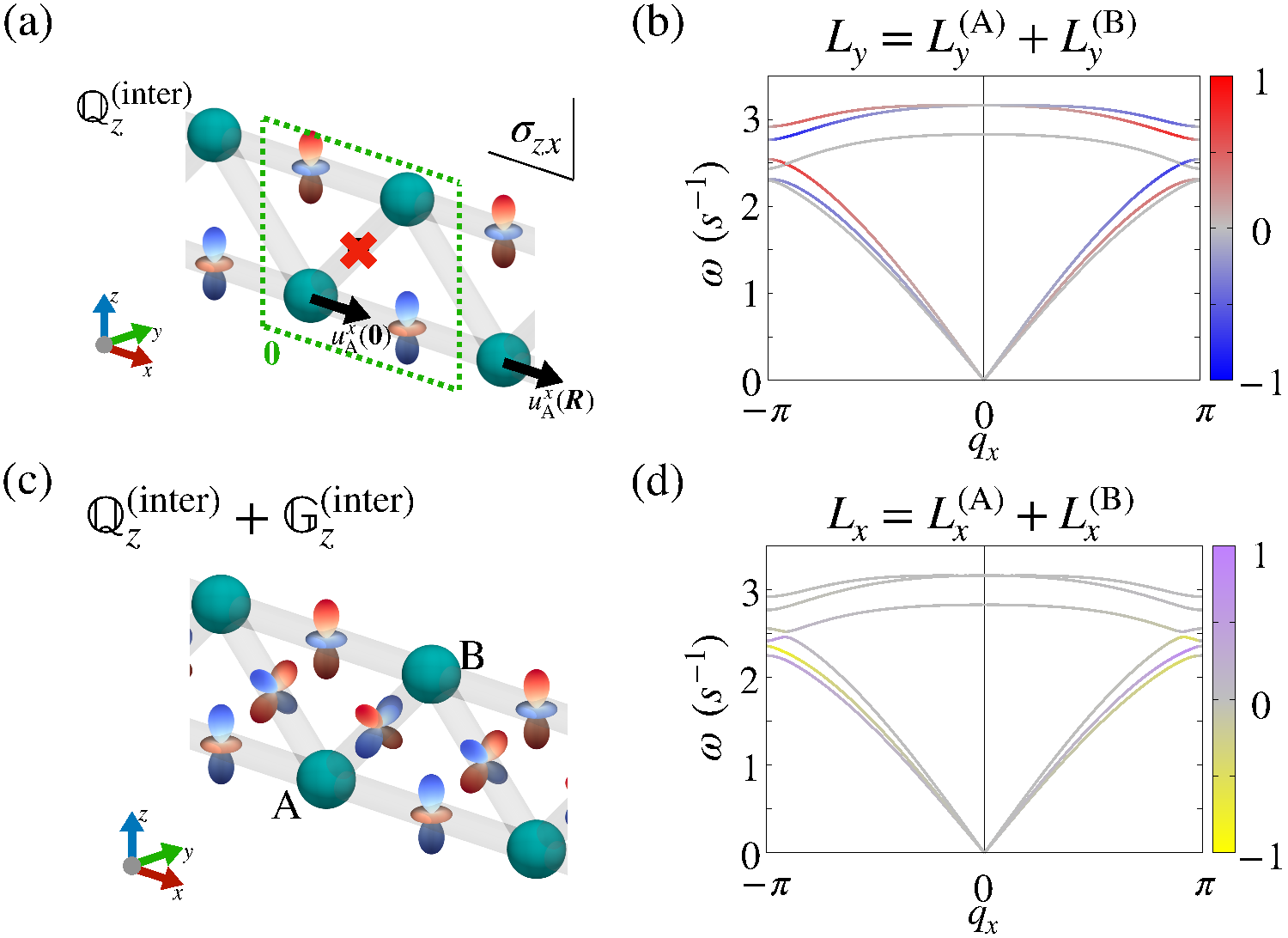}
\caption{\justifying
(a) Schematic representation of the electric dipole $\hat{\mathbb{Q}}_{z}^{\rm (inter)}$ defined in Eq.~(\ref{eq_Qz}), and (b) the corresponding phonon energy dispersion with the phonon AM distribution.
(c) Combined order of $\hat{\mathbb{Q}}_{z}^{\rm (inter)}$ and $\hat{\mathbb{G}}_{z}^{\rm (inter)}$ and (d) the corresponding phonon energy dispersion with phonon AM distribution.
}
\label{fig_qz_chiral}
\end{figure}

Next, we turn to a polar SAMB, in place of the ferroaxial one, corresponding to an electric polarization along the $z$ direction. 
Within the SAMB framework, such a contribution is incorporated as
\begin{align}
\hat{\mathbb{Q}}_{z}^{\rm (inter)} = \hat{Q}_{u}^{\rm (a)} \otimes \hat{Q}_{z}^{\rm (b)},
\label{eq_Qz}
\end{align}
which belongs to the $\mathrm{B}_{1u}$ irrep.
Here, the bond-cluster component satisfies $-[\hat{Q}_{z}^{\rm (b)}]_{\rm AA}(\pm\bm{R}) = [\hat{Q}_{z}^{\rm (b)}]_{\rm BB}(\pm\bm{R}) = 1/2$. 
This term breaks $\mathcal{P}$ symmetry as well as the mirror symmetry $\sigma_{xy}$, as shown in Fig.~\ref{fig_qz_chiral}(a). 
The corresponding phonon dispersion exhibits a Rashba-type antisymmetric splitting of the phonon AM, characterized by $L_{y}^{\rm (ph)}(q_x) = -L_{y}^{\rm (ph)}(-q_x)$, as shown in Fig.~\ref{fig_qz_chiral}(b).

We then consider the combined effect of the ferroaxial and polar SAMBs. 
As shown in Fig.~\ref{fig_qz_chiral}(c), introducing both $\hat{\mathbb{G}}_{z}^{\rm (inter)}$ and $\hat{\mathbb{Q}}_{z}^{\rm (inter)}$ breaks all mirror symmetries as well as $\mathcal{P}$ symmetry, thereby realizing a chiral system. 
In the ferroaxial state alone, the sublattice-resolved chiral components $G_u^{\pm}$ cancel due to the remaining mirror symmetry $\sigma_{xy}$. 
However, once the polar term is introduced, this symmetry is lifted, rendering the A and B sublattices inequivalent. 
As a result, the cancellation of local chirality is removed, and a finite global chirality emerges. 
This is clearly reflected in the phonon AM distribution shown in Fig.~\ref{fig_qz_chiral}(d), where a net antisymmetric component of $L_x^{\rm (ph)}$ develops along $q_x$. 

Finally, we discuss the minimal microscopic model parameters responsible for the characteristic phonon AM splitting. 
For the ferroaxial case, the hidden sublattice-resolved splitting $L_x^{({\rm ph}),({\rm A,B})}$ already appears with the NN isotropic force constant $k^{(1)}$ in $U$ given by Eq.~(\ref{eq_potential}).
In contrast, the polarization-induced splitting $L_y^{\rm (ph)}$ requires both $k^{(1)}$ and $k^{(1)'}$, reflecting the necessity of additional NN off-diagonal component.
When both ferroaxiality and polarity are present, a chiral phonon AM splitting emerges with $k^{(1)}$ alone, without requiring additional off-diagonal force constants.

%%%%%%%%%%%%%
%%%   Conclusion  %%%
%%%%%%%%%%%%%

We developed a symmetry-based framework for constructing effective harmonic phonon models using the SAMB.
By decomposing the force-constant matrix under symmetry constraints and the ASR, we identified the minimal ingredients responsible for phonon AM splitting.
Applying this framework to a minimal zigzag-chain {model}, we show that a ferroaxial SAMB generates hidden sublattice-resolved phonon AM that cancels globally
due to $\mathcal{P}$ and $\mathcal{T}$ symmetries.
Introducing a polar SAMB lifts this cancellation by breaking inversion and mirror symmetries, leading to a chiral phonon dispersion. 
Thus, the interplay of ferroaxiality and polarity converts hidden chirality into observable phonon responses.
Our results provide a unified symmetry-adapted description of phonon AM splitting across ferroaxial, polar, and chiral systems, and offer a systematic route to design phononic responses via symmetry breaking.

{
\begin{acknowledgments}
The authors thank H. Kusunose for fruitful discussions.
{This research was supported by JSPS KAKENHI Grants Numbers JP22H00101, JP23H04869, and by JST CREST (JPMJCR23O4) and JST FOREST (JPMJFR2366).} 
\end{acknowledgments}
}
\bibliographystyle{plain}
\bibliography{refs}

\end{document}